\documentclass[twocolumn,aps,showpacs,superscriptaddress]{revtex4}

\usepackage{amsmath,amssymb}
\usepackage[]{graphicx}
\usepackage{dcolumn}
\usepackage{bm}
\usepackage[dvips]{color} 

\input{MyMc}

\begin{document}

\title{%
  Thermal escape of fractional vortices in long Josephson junctions
}

\author{U.~Kienzle}
\author{T.~Gaber}
\author{K.~Buckenmaier}
\affiliation{%
  Physikalisches Institut -- Experimentalphysik II and Center for Collective Quantum Phenomena,
  Universit\"{a}t T\"{u}bingen,
  Auf der Morgenstelle 14,
  D-72076 T\"{u}bingen, Germany
}%

\author{K.~Ilin}
\author{M.~Siegel}
\affiliation{%
  Universit\"{a}t Karlsruhe,
  Institut f\"{u}r Mikro- und Nanoelektronische Systeme,
  Hertzstra\ss e 16,
  D-76187 Karlsruhe, Germany
}%

\author{D.~Koelle}%
\author{R.~Kleiner}%
\author{E.~Goldobin}
\affiliation{%
  Physikalisches Institut -- Experimentalphysik II and Center for Collective Quantum Phenomena,
  Universit\"{a}t T\"{u}bingen,
  Auf der Morgenstelle 14,
  D-72076 T\"{u}bingen, Germany
}%


\begin{abstract}

   We consider a fractional Josephson vortex in a long 0-$\kappa$ Josephson junction. A uniformly applied bias current exerts a Lorentz force on the vortex. If the bias current exceeds the critical current, an integer fluxon is torn off the $\kappa$-vortex and the junction switches to the voltage state.
   In the presence of thermal fluctuations the escape process takes place with finite probability already at subcritical values of the bias current.
   We experimentally investigate the thermally induced escape of a fractional vortex by high resolution measurements of the critical current as a function of the topological charge $\kappa$ of the vortex and compare the results to numerical simulations for finite junction lengths and to theoretical predictions for infinite junction lengths. To study the effect caused by the junction geometry we compare the vortex escape in annular and linear junctions.

\end{abstract}


\date{\today}
\maketitle


\section{Introduction}
 \label{Sec:Introduction}
Vortices in long Josephson junctions (LJJs) usually carry a single magnetic flux quantum $\Phi_0$ and therefore are called fluxons. The study of fluxons has been attracting a lot of attention during the last 40 years because of their interesting nonlinear nature \cite{Barone:PhysicsJE, Likharev:DynamicsJJ, Ustinov98} as well as because of possible applications \cite{Koshelets2000, Kemp02, Ozyuzer07, Beck05}.

Recently it was shown that one can create and study experimentally vortices that carry only a fraction of the magnetic flux quantum. Initially, vortices carrying only half of the magnetic flux quantum, so-called semifluxons \cite{Bulaevskii78, Goldobin02, Xu95}, were observed and studied.
They exist in 0-$\pi$ LJJs, which can be fabricated using superconductors with an anisotropic order parameter that changes sign depending on the direction in $k$ space (\eg, $d$-wave order parameter symmetry \cite{Hilgenkamp03, Kirtley05, Smilde02, VanHarlingen95}) or with an oscillating order parameter (\eg, with a ferromagnetic barrier \cite{Weides06b}).

In 0-$\pi$ LJJs, some parts behave as 0 junctions and other parts as $\pi$ junctions. In this kind of junctions, the ground state phase $\mu(x)$ will have the value 0 deep inside the 0-region and the value $\pi$ deep inside the $\pi$-region. In the $\lambda_J$-vicinity of the 0-$\pi$ boundary, $\mu(x)$ will continuously change from 0 to $\pi$, where $\lambda_J$ is the Josephson penetration depth. Because of the phase bending a magnetic field $\propto d\mu/dx$ localized in the $\lambda_J$-vicinity of the boundary appears and a supercurrent $\propto \sin[\mu(x)]$ circulates around it \cite{Bulaevskii78, Goldobin02, Xu95}. The total magnetic flux localized at the 0-$\pi$ boundary is equal to $+\Phi_0/2$. Since the Josephson phase is $2\pi$-periodic, its value can be $-\pi$ inside the $\pi$-region instead of $\pi$. In this case, the supercurrent flows counterclockwise and an antisemifluxon with the magnetic flux $-\Phi_0 /2$ is formed.

It turns out that instead of a $\pi$-discontinuity of the Josephson phase at the 0-$\pi$ boundary one can artificially create any arbitrary $\kappa$-discontinuity at any point of the LJJ and the value of $\kappa$ can be tuned electronically \cite{Goldobin04c}. At the resulting 0-$\kappa$ boundary two types of vortices may exist \cite{Goldobin04a}: a direct vortex with topological charge $-\kappa$ ($-\kappa$ vortex) and a complementary vortex with topological charge $2\pi-\kappa$ ($2\pi-\kappa$ vortex). Without loosing generality we assume that $0<\kappa<2\pi$. Such vortices are a generalization of semifluxons and antisemifluxons discussed above. Only if their topological charge is less or equal than $2\pi$ by absolute value, they are stable \cite{Goldobin05a}.

By applying a spatially uniform bias current to a LJJ with a $\kappa$ vortex, one exerts a Lorentz force which pushes the vortex along the junction. Depending on the mutual polarity of the vortex and the bias current, the direction of the force can be adjusted. Since the vortex exists to compensate the phase discontinuity, it is pinned at it, \ie, it may only bend under the action of the Lorentz force, but it cannot move away from the discontinuity. However, increasing the bias current,  the force can be made strong enough to tear off a whole integer fluxon out of a $\kappa$ vortex. The fluxon moves away along the junction, leaving a complementary $(\kappa-2\pi)$ vortex pinned at the discontinuity. Further dynamics leads to the switching of the 0-$\kappa$ LJJ into the voltage state. In annular LJJs, this process takes place when the normalized bias current $\gamma=I/I_{{\rm c0}}$ reaches the critical current of \cite{Goldobin04b, Malomed04}
\begin{equation}
  \gamma_c(\kappa)=\left|{\frac{\sin(\kappa/2)}{\kappa/2}}\right|
  , \label{Eq:depinning current}
\end{equation}
where $I_{{\rm c0}}=j_cwL$ is the ``intrinsic'' critical current, which corresponds to the measurable critical current if $\kappa=0$; $w$ is the LJJ width and $L$ is its length.
Due to thermal fluctuations, the escape process described above will already take place with finite probability at $\gamma\lesssim\gamma_c$. In this paper, we study the thermal escape of an arbitrary $\kappa$ vortex experimentally and compare it to numerical simulations for finite junction lengths and to theoretical predictions for infinite lengths.

The paper is organized as follows. In Sec.~\ref{Sec:Point-like JJ}, we briefly review the thermal escape of the phase in a point-like Josephson junction. Then, in Sec.~\ref{Sec:long junctions with a fractional vortex}, we discuss theoretically the phase escape in a long JJ with a vortex and compare it with our experimental findings in Sec.~\ref{Sec:Experiments}. The influence on junction geometry is further investigated in Sec.~\ref{Sec:Linear junctions}.
\section{Point-like Josephson junction}
 \label{Sec:Point-like JJ}
\begin{figure}[!htb]
  \begin{center}
    \includegraphics{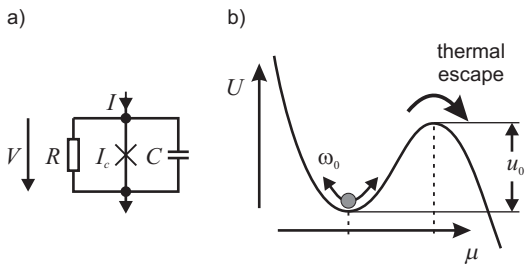}
  \end{center}
  \caption{%
    (a) Resistively and capacitively shunted junction (RCSJ) model. (b) Particle in a tilted washboard potential. The thermal escape from the metastable state is indicated by an arrow.
  }
  \label{Fig:RCSJ and metastable state}
\end{figure}

In the Stewart\--McCumber model \cite{Stewart68, McCumber68}, see Fig.~\ref{Fig:RCSJ and metastable state}(a), the dynamics of a current-biased point-like JJ is described by an equation of motion for the Josephson phase\:$\mu$
\begin{equation}
  C\left(\frac{\Phi_0}{2\pi}\right)\ddot{\mu}+\frac{1}{R}\left(\frac{\Phi_0}{2\pi}\right)\dot{\mu}+I_c\sin(\mu)-I=0
  , \label{Eq:equation of motion}
\end{equation}
where $C$ is the effective shunt capacitance, $R$ is the effective shunt resistance of the JJ embedded in the bias circuitry and dots denote the derivative with respect to time.
The equation of motion \eqref{Eq:equation of motion} is equivalent to the damped motion of a particle of mass $m=C(\Phi_0/2\pi)$ along the generalized coordinate $\mu$ in a tilted washboard potential
\begin{equation}
U(\mu)=-E_J(\gamma\mu+\cos\mu)
  , \label{Eq:potential}
\end{equation}
where $E_J=\Phi_0I_c/2\pi$ is the Josephson coupling energy, see Fig.~\ref{Fig:RCSJ and metastable state}(b). The tilting angle of the potential is proportional to the bias current $\gamma=I/I_c$ applied.

The process of the particle escape from a metastable minimum in this Kramers-like system \cite{Kramers40, Haenggi90} corresponds to a transition of the JJ from a superconducting zero-voltage state to a finite voltage state.
In the absence of thermal and quantum fluctuations the junction switches from the zero-voltage state, which corresponds to the particle being localized in one of the potential wells, to a finite voltage state, when the minima disappear and the particle runs down the potential, at $\gamma\geqslant1$. If, like in our case, the temperature is finite, the particle may escape from the well already at $\gamma\lesssim1$, beeing thermally excited over the barrier \cite{Kramers40, Haenggi90, Vion96, Fulton74, Castellano96, Silvestrini88b, Turlot89, Washburn84}.

The rate at which this process occurs depends on the barrier height
\begin{equation}
u_0=2E_J\left[\sqrt{1-\gamma^2}-\gamma\arccos(\gamma)\right]
   \label{Eq:barrier height}
\end{equation}
and the small amplitude oscillation frequency of the particle at the bottom of the well,
\begin{equation}
\omega_0=\sqrt{\frac{\partial^2U/\partial\mu^2}{m}}=\omega_p\left(1-\gamma^2\right)^{1/4}
  . \label{Eq:oscillation frequency}
\end{equation}
Here, $\omega_p=\sqrt{2\pi I_c/\Phi_0 C}$ is the plasma frequency. For $1-\gamma\ll 1$, Eq.~\eqref{Eq:barrier height} can be approximated as
\begin{equation}
u_0\approx E_J\frac{4\sqrt{2}}{3}\left(1-\gamma\right)^{3/2}
  . \label{Eq:approximated barrier height}
\end{equation}
In the thermal regime the escape of the particle from the well occurs at a bias-current dependent rate of \cite{Kramers40, Haenggi90}
\begin{equation}
\Gamma(\gamma)=a\frac{\omega_0(\gamma)}{2\pi}\exp\left[-\frac{u_0(\gamma)}{k_BT}\right]
  , \label{Eq:escaperate}
\end{equation}
where $a\lesssim 1$ is a damping dependent prefactor.
Knowing all relevant junction parameters and the bath temperature $T$, one can calculate the escape rate $\Gamma$ of a current-biased point-like JJ.
\section{long Josephson junction with a fractional vortex}
 \label{Sec:long junctions with a fractional vortex}
\begin{figure}[!htb]
  \begin{center}
    \includegraphics{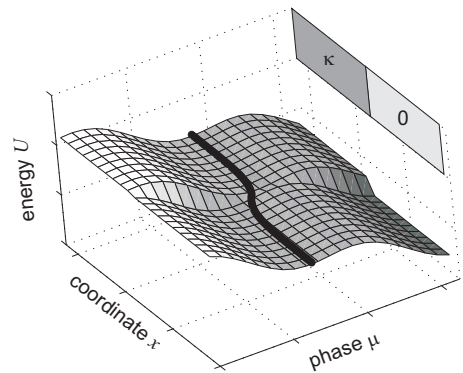}
  \end{center}
  \caption{%
    LJJ in the picture of a particle chain in an extended potential.
  }
  \label{Fig:washboard}
\end{figure}

Here, the considered system, a long JJ with a fractional vortex, is more complex than a point-like JJ. In order to account for the spatial phase variation $\mu(x)$, one has to extend the picture of the particle in a tilted washboard potential. First, to take into account the finite junction length, one replaces the point-like particle in a 1D-potential by an elastic string in an extended potential. Second, the $\kappa$ vortex in the LJJ corresponds to a kink in the string, as illustrated in Fig.~\ref{Fig:washboard}.

The depinning process of a fractional vortex at $\gamma\lesssim\gamma_c(\kappa)$ [see Eq.~\eqref{Eq:depinning current}], described above, corresponds to the escape of this string out of a metastable well. To find the effective barrier height for this system, we map the multidimensional system to a point-like particle in an effective 1D potential. By comparing the escape rate $\Gamma$, which we obtain by high resolution measurement of the depinning current, to the known escape rate of a point-like JJ, one can make a statement about the effective barrier height.

For an \emph{infinitely long} JJ, the effective potential for the escape of a fractional vortex close to the depinning current can be derived analytically, as shown in Ref.~\cite{Vogel08}. Here, only the main results are summarized. The barrier height is given by
\begin{equation}
u_v = E'_J\frac43(\kappa\gamma_c)^{3/2} (3F)^{-1/2}\left(1-\frac{\gamma}{\gamma_c}\right)^{3/2},
\label{Eq:DU2}
\end{equation}
with
\begin{eqnarray}
F & = & \frac23\sin\frac{\kappa}{2}\left(\sin\beta_c-\gamma_c\beta_c\right),\\
\beta_c & = & \frac{\pi}{2}-\arcsin(\gamma_c),
\label{Eq:F and BetaC}
\end{eqnarray}
and the depinning current $\gamma_c(\kappa)$ given by Eq.~\eqref{Eq:depinning current}. $E'_J=\Phi_0 j_c w\lambda_J/2\pi=E_J\lambda_J/L$ is the Josephson coupling energy per normalized unit length and $j_c$ the junction's critical current density assumed to be equal in 0 and $\kappa$ parts. Here, we use $u_v$ and reserve $u_0$ as the barrier height for the homogeneous phase escape (short junction limit).
The small oscillation frequency \eqref{Eq:oscillation frequency} in this case is given by the eigenfrequency of the fractional vortex. In absence of a bias current the eigenfrequency is given by \cite{Goldobin05a, Buckenmaier07}
\begin{equation}
  \omega_0(\kappa)=\omega_p\sqrt{%
    \frac{1}{2}\cos\frac{\kappa}{4}\left(\cos\frac{\kappa}{4}+\sqrt{4-3\cos^2\frac{\kappa}{4}}\right)
  }.
  \label{Eq:eigenfrequency}
\end{equation}
For $\gamma\neq0$ the analytical expression for $\omega_0(\kappa,\gamma)$ is unknown, but can be approximated as 
\begin{equation}
  \omega_0(\kappa,\gamma)\approx
  \omega_0(\kappa,0)\sqrt[4]{1-\left(\frac{\gamma}{\gamma_c(\kappa)}\right)^2}.
  \label{Eq:eigenfrequency,gamma}
\end{equation}

For LJJs of \emph{finite length} one may use a numerical approach and follow the ansatz described by Castellano \etal\cite{Castellano96}. As in a point-like JJ the thermal activation occurs over a barrier that is defined by the energy difference between a minimum and the lowest adjacent saddle point/maximum of the potential surface. The corresponding phase distributions are solutions of the stationary sine-Gordon equation 
\begin{equation}
\mu_{xx}-\sin\left[\mu+\kappa H(x)\right]+\gamma=0
  , \label{Eq:sine-Gordon}
\end{equation}
where $H(x)$ is a step function that describes the position of the 0 and $\kappa$ region
\begin{equation}
 H(x)=
  \begin{cases}
  0 & \text{for $x<0$}\\
  1 & \text{for $x>0$.}
  \end{cases}
  \label{Eq:step function}
\end{equation}
For a linear junction geometry at zero applied magnetic field the phase obeys the boundary conditions 
\begin{equation}
\mu_x\left(-\frac l2\right)=\mu_x\left(\frac l2\right)=0.
\label{Eq:boundary conditions}
\end{equation}
Here, $\mu(x)$ is the Josephson phase, which is a continuous function of $x$, the spatial coordinate $x$ is normalized to the Josephson penetration depth $\lambda_J$ and $l=L/\lambda_J$ is the normalized junction length. The subscript $x$ denotes the derivate with respect to coordinate $x$.
The potential energy of any state $\mu(x)$ is given by 
\begin{equation}
U=E_J \frac1l \int_{-l/2}^{l/2}\left\{1-\cos[\mu+\kappa H(x)]-\gamma\mu + \frac12\left(\mu_x\right)^2\right\} dx.
\label{Eq:Pot Energy}
\end{equation}
Let us denote the solutions corresponding to energy minima as $\mu_a(x)$ and the ones corresponding to saddle points/maxima as $\mu_s(x)$. Then, the effective barrier height is 
\begin{equation}
u=U(\mu_s)-U(\mu_a)
  , \label{Eq:potential difference u}
\end{equation}
which is defined for each pair of $\mu_s$, $\mu_a$.

For an annular geometry it is more convenient to rewrite and solve the sine-Gordon equation~\eqref{Eq:sine-Gordon} for the (discontinuous) phase $\phi(x)=\mu(x)+\kappa H(x)$:
\begin{equation}
\phi_{xx}-\sin\phi+\gamma+\kappa\delta_x=0
  , \label{Eq:sine-Gordon Phi}
\end{equation}
where $\delta_x$ is the derivate of the $\delta$-function. The vortex is placed at $x=0$, which in annular geometry corresponds to $x=l$, and the following boundary conditions are used
\begin{equation}
\begin{array}{rcl}
\phi(0)&=&\phi(l)+\kappa,\\
\phi_x(0)&=&\phi_x(l).
\end{array}
   \label{Eq:boundary conditions2}
\end{equation}
If $\kappa>0$, then the topological charge of the vortex is $-\kappa<0$.
Note, that all static solutions of Eqs.~\eqref{Eq:sine-Gordon Phi} and \eqref{Eq:boundary conditions2} also obey \cite{Malomed04}
\begin{equation}
\gamma=\kappa^{-1}\left[(1-\cos\kappa)\cos\phi(0)-\sin\kappa\sin\phi(0)\right]
  . \label{Eq:static solutions}
\end{equation}
Here, $\phi(0)$ denotes the phase value at $x=0$. By inverting the signs of $\kappa$ and $\phi(0)$, Eq.~\eqref{Eq:sine-Gordon Phi} and Eq.~\eqref{Eq:boundary conditions2} can also be applied to a vortex with positive charge. Eq.~\eqref{Eq:static solutions} is particularly useful, if a shooting algorithm is used to find solutions of Eqs.~\eqref{Eq:sine-Gordon Phi} and \eqref{Eq:boundary conditions2}.

For $\kappa=0$, the effect of junction length (and an externally applied magnetic field) on the phase escape in \emph{linear} LJJs has already been discussed by other authors \cite{Castellano96}. In the following we will summarize some of the results, particularly important for the situation discussed in this paper. That is, we do not consider an applied magnetic field. In short junctions, $l<\pi$, the barrier height $u$ is identical to $u_0$ for any $\gamma$, \ie, the value obtained from the short junction limit formula~\eqref{Eq:barrier height}. The phase configurations corresponding to minima and saddle points are homogeneous, \ie, $\mu(x)=const.$. In longer junctions, $l>\pi$, phase escape via additional, inhomogeneous saddle solutions of lower potential barrier \eqref{Eq:potential difference u} results in $u\leq u_0$. This is easily understood, as the energy required for a homogeneous phase escape increases linearly with junction length, whereas it is constant\footnote{The necessary energy is $u=8E_J/l$ for $\gamma=0$.} for the alternative process, that is the creation of a fluxon/antifluxon and its successive propagation along the junction. However, the nature and ultimate existence of the saddle solutions depends on bias current $\gamma$. As it turns out, at dc-bias close to the critical current, $\gamma_c-\gamma\ll 1$, which is of primary relevance in experiment, usually only one minimum and one saddle point solution remains. For junction lengths $l<9$ and bias currents $\gamma\sim0.99\gamma_c$, the barrier heights $u$ even coincide with the short junction value $u_0$.
In \emph{annular} junctions the length at which the phase escapes via inhomogeneous phase configurations is $2\pi$ instead of $\pi$, as boundary conditions require the creation of a fluxon-antifluxon pair. As in linear junctions, the barrier height $u$ coincides with $u_0$ for $\gamma\sim0.99\gamma_c$ and $l\leq12$, which is the longest junction considered here.

\begin{figure}[!htb]
  \begin{center}
    \includegraphics{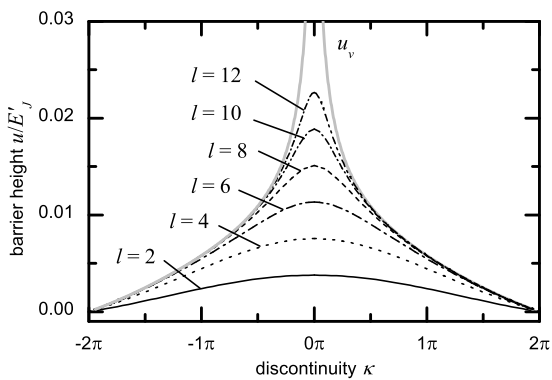}
  \end{center}
  \caption{%
    Effective barrier height at $\gamma=0.99\gamma_c$ as a function of $\kappa$ for different junction lengths. The gray line corresponds to the analytic result from Eq.~\eqref{Eq:DU2}.
  }
  \label{Fig:Simu annular}
\end{figure}

Now, how does the presence of a fractional vortex, \ie, $|\kappa|>0$, affect the phase escape? In the following we first consider an annular geometry, where junction boundaries do not play a role. The influence of open boundaries on phase escape is discussed in Sec.~\ref{Sec:Linear junctions}. Fig.~\ref{Fig:Simu annular} compares the effective barrier height $u(\kappa)$ at $\gamma=0.99\gamma_c(\kappa)$ for different junction lengths. 
Here, $u$ is normalized to $E'_J$ to avoid a normalization that depends on JJ length. The analytic result $u_v(\kappa)$ for $l=\infty$ is also depicted in the graph. As can be seen, the presence of a vortex facilitates the phase escape by lowering the effective barrier height. The larger the vortex, the easier is the escape. Fig.~\ref{Fig:Simu annular} also shows, that with increasing junction length, $u(\kappa)$ asymptotically approaches $u_v(\kappa)$ -- the analytic result for an infinitely long junction.

For further discussion it is important to note, that although $u$ strongly depends on bias current, numerical calculations of $u(\gamma$) have shown, that it scales as $(1-\gamma/\gamma_c)^{3/2}$ close to critical current with a pre-factor almost independent of $\gamma$. This is very convenient, as $u$, $u_v$ [compare Eq.~\eqref{Eq:DU2}] and $u_0$ [compare Eq.~\eqref{Eq:approximated barrier height}] exhibit the same asymptotic behavior for $\gamma\rightarrow\gamma_c$. With their relative proportions independent of $\gamma$, they can be easily compared with experimental data. 
\section{Experiments}
 \label{Sec:Experiments}
For the experiments we used tunnel Nb-Al-AlO$_{x}$-Nb annular LJJs (ALJJ) with different junction radii $R$. The junction properties are specified in Tab.~\ref{Tab:ALJJ}, the specific capacitance of all junctions is $C'\simeq 4.1 \mu{\rm F/cm^{2}}$. The Josephson penetration depth $\lambda_J$ was estimated taking into account the idle region \cite{Monaco95}. The normalized length of the ALJJ is given by $l=2\pi R/\lambda_J$.
\begin{table}[!htb]
\centering
\begin{tabular}{|l|l|l|l|l|l|}
\hline
\multicolumn{1}{|c|}{$R$} & \multicolumn{1}{c|}{$w$} & \multicolumn{1}{c|}{$j_c$} & \multicolumn{1}{c|}{$\lambda_J$} &      \multicolumn{1}{c|}{$l$} & \multicolumn{1}{c|}{$I^{{\rm min}}_{{\rm inj}}$} \\ 
\multicolumn{1}{|c|}{[$\mu$m]} & \multicolumn{1}{c|}{[$\mu$m]} & \multicolumn{1}{c|}{[A/cm$^2$]} & \multicolumn{1}{c|}{[$\mu$m]} & \multicolumn{1}{c|}{} & \multicolumn{1}{c|}{[mA]} \\ 
\hline
\multicolumn{1}{|c|}{30} & \multicolumn{1}{c|}{5} & \multicolumn{1}{c|}{134} & \multicolumn{1}{c|}{37.9} & \multicolumn{1}{c|}{5.0} & \multicolumn{1}{c|}{7.4} \\ 
\hline
\multicolumn{1}{|c|}{50} & \multicolumn{1}{c|}{5} & \multicolumn{1}{c|}{87.8} & \multicolumn{1}{c|}{46.7} & \multicolumn{1}{c|}{6.7} & \multicolumn{1}{c|}{7.0} \\ 
\hline
\multicolumn{1}{|c|}{70} & \multicolumn{1}{c|}{5} & \multicolumn{1}{c|}{72.6} & \multicolumn{1}{c|}{51.4} & \multicolumn{1}{c|}{8.6} & \multicolumn{1}{c|}{6.5} \\ 
\hline
\end{tabular}
\caption{Junction parameters (annular) at $T=4.2\,{\rm K}$.}
\label{Tab:ALJJ}
\end{table}
\begin{figure}[!htb]
  \begin{center}
    \includegraphics{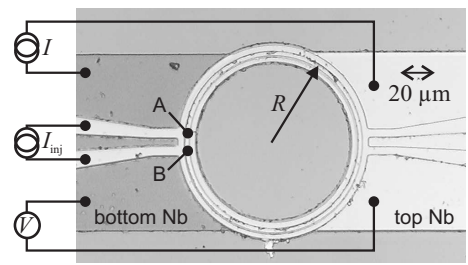}
  \end{center}
  \caption{%
    Optical image (top view) of one of the investigated samples: ALJJ with two pairs of injectors. The right pair of current injectors was not used during experiment. 
  }
  \label{Fig:sample}
\end{figure}

\begin{figure}[!htb]
  \begin{center}
    \includegraphics{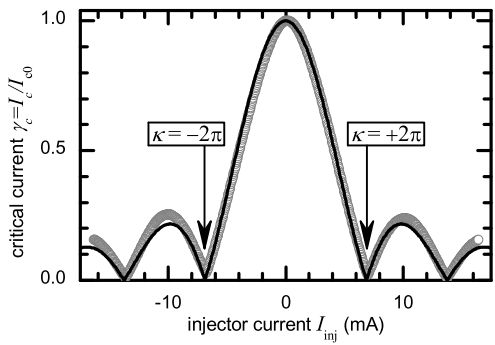}
  \end{center}
  \caption{%
    The dependence $\gamma_c (I_{{\rm inj}})$ measured at $T\approx 4.2\,{\rm K}$ (gray symbols) and the corresponding theoretical curve $\gamma_c (\kappa)$ (continuous line).
  }
  \label{Fig:Fraunhofer}
\end{figure}

To create an arbitrary $\kappa$ discontinuity in the junctions, a pair of tiny current injectors is used, as shown in Fig.~\ref{Fig:sample}. The short section of the top electrode between points A and B of a length $\ll \lambda_J$ has an inductance $L_{{\rm inj}}$. A current $I_{{\rm inj}}$ passing through this inductance creates a phase drop $\kappa=L_{{\rm inj}}I_{{\rm inj}}2\pi/\Phi_0$ across the distance AB, \ie, the $\kappa$ discontinuity. To calibrate the injectors we have measured the critical current $I_c$ as a function of $I_{{\rm inj}}$ for the left injector pair in Fig.~\ref{Fig:sample}. The $I_c\left(I_{{\rm inj}}\right)$ pattern is presented for the shortest junction with $R=30\,\mu{\rm m}$ in Fig.~\ref{Fig:Fraunhofer} and looks like a perfect Fraunhofer pattern in accordance to the theory \cite{Goldobin04b, Nappi02, Malomed04}. The first minimum at $I^{{\rm min}}_{{\rm inj}}$ corresponds to $\kappa=2\pi$. The values $I^{{\rm min}}_{{\rm inj}}$ are given in Tab.~\ref{Tab:ALJJ}. Thus, for any $I_{{\rm inj}}$, the corresponding value of $\kappa$ can be calculated as $\kappa=2\pi I_{{\rm inj}}/\left|I^{{\rm min}}_{{\rm inj}}\right|$.

The escape of a fractional vortex in an ALJJ is experimentally investigated by measuring the statistics of switching of the junction from the zero-voltage state to a finite voltage state. To do so, a bias current applied to the junction is ramped up at a constant rate $\dot{I}$ and the current $I_c$ at which the junction switches from its zero-voltage state is recorded \cite{Fulton74}. The probability distribution $P(I)$ of such switching currents is found by accumulating a large number of measurements of $I_c$ and generating a histogram. Using the obtained $P(I)$ distribution, the bias current dependent escape rate can be reconstructed as \cite{Fulton74}
\begin{equation}
\Gamma(I)=\frac{\dot{I}}{\Delta I}\ln \frac{\int^{\infty}_{I}{P(I')dI'}}{\int^{\infty}_{I+\Delta I}{P(I')dI'}}
  , \label{Eq:escaperate experimentally}
\end{equation}
where $\Delta I$ is the width of the bins on the histogram.
The shape of the histogram depends on the bias current ramp rate, the bath temperature $T$ and damping \cite{Silvestrini88a}. Higher $T$ leads to a broader histogram shifted towards lower currents. Note, electronic noise in the measurement setup has, to the first order, the same effect as the increase of $T$. Since the accuracy of our measurement is defined by the width of the escape histogram, special measures were taken to suppress electronic noise down to the level where the broadening caused by electronic noise is well below the one caused by the bath temperature. We mounted the sample in a copper box to shield it from electromagnetic radiation and in a cryoperm cylinder to shield it from static magnetic fields. The bias lines contained three-stage low pass filters: two cold filters in close vicinity of the sample and one at room temperature. The measurement technique and setup are similar to those described in Ref.~\cite{Wallraff03b}.

Following the analysis method used in Refs.~\cite{Wallraff03b, Castellano96, Fulton74}, we fit the escape rate $\Gamma(I)$ calculated from the measured histogram using Eq.~\eqref{Eq:escaperate experimentally} with the known escape rate of a point-like JJ, see Eq.~\eqref{Eq:escaperate}, having the relative barrier height $u/u_0$ and $\gamma_c I_{{\rm c0}}$ as fitting parameters. The damping dependent prefactor was assumed to be $a=1$.
\begin{figure}[!htb]
  \begin{center}
    \includegraphics{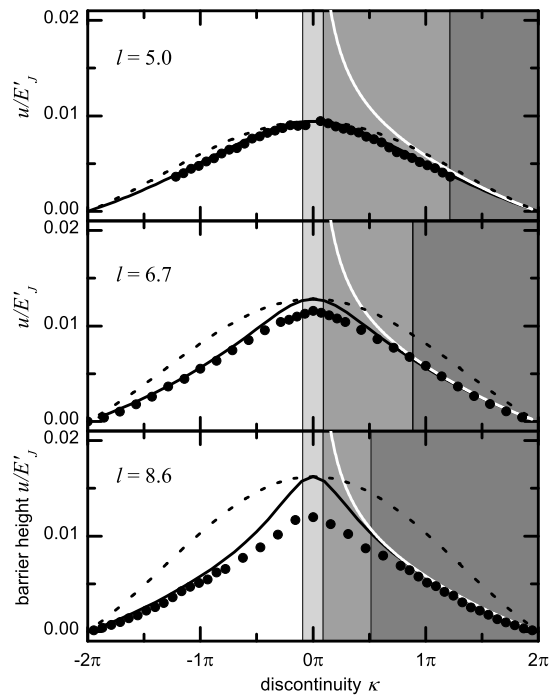}
  \end{center}
  \caption{%
Effective barrier height $u(\kappa)$ for different junction lengths $l$ at $T=4.2\,{\rm K}$. Solid circles correspond to experimental results, white lines to theoretical predictions for infinite 0-$\kappa$ LJJ and solid lines to numeric simulations using the sine-Gordon equation in ALJJ of given length. Dashed lines show barrier heights for point-like junctions with rescaled $\gamma_c$. 
  }
  \label{Fig:experiment annular}
\end{figure}
 
The effective barrier height found in this way is shown in Fig.~\ref{Fig:experiment annular} for 3 different junction lengths at a bath temperature $T=4.2\,\mathrm{K}$. Solid circles indicate experimental results and the black solid line corresponds to numerical simulations. For all junction lengths, the experimental data are in good agreement with simulations in particular for large $\kappa$ values. In the range of small values of the discontinuity $\kappa$, the deviation from simulations increases with junction length. There are two possible reasons explaining the deviations.

First, this might be due to junction inhomogeneities. For large $|\kappa|$, only the $\lambda_J$-vicinity of the $\kappa$-discontinuity is important for the escape, whereas for $|\kappa|\to 0$ the whole junction matters. The longer the junction is, the higher is the probability for inhomogeneities, thus the deviations are more dominant for longer JJs.

Second, our calculations are based on a single mode approximation, which becomes less accurate and even fails in the limit $|\kappa| \to 0$ and $l\to\infty$\cite{Vogel08}. This happens because the eigenfrequency of the lowest eigenmode $\omega_0$ and the next ones $\omega_1$, $\omega_2$ etc. are not well separated at the values of bias current where the escape takes place. At first, it seems that at $\gamma\to\gamma_c$ the eigenfrequency $\omega_0\to0$, while $\omega_1(\gamma_c)$ stays finite. Therefore there is always a region around $\gamma_c$ where $\omega_0\ll\omega_1$, regardless how small is the separation between $\omega_0$ and $\omega_1$ at low $\gamma$, say, at $\gamma=0$. However, the peak of the escape histogram (even in the case of quantum escape with the rightmost position of the histogram) is not sufficiently close to $\gamma_c$ in order to keep the condition $\omega_0\ll\omega_1$ satisfied. Then it is important how much $\omega_0$ and $\omega_1$ are separated at those values of $\gamma$ that are most relevant for the escape process. This separation is roughly proportional to, but much smaller than $\omega_1(0)-\omega_0(0)$. In the limit $l \to \infty$ and $|\kappa| \to 0$, according to Eq.~\eqref{Eq:eigenfrequency}, $\omega_0(0) \to \omega_p$, while the the next mode (the lowest mode in the plasma band) $\omega_1=\omega_p$. Thus, the single mode approximation fails. To make it working again one has either to increase $|\kappa|$ or to reduce $l$. By increasing $|\kappa|$ one shifts $\omega_0(\kappa,0)$ downwards from $\omega_1=\omega_p$, according to Eq.~\eqref{Eq:eigenfrequency}. By decreasing $l$ one increases the wave vector $k_1={2\pi}/{l}$ of the lowest mode in the plasma band, thus, shifting $\omega_1(0)$ up as
\[
  \omega_1(0)=\omega_p\sqrt{1+k_1^2}=\omega_p\sqrt{1+({2\pi}/{l})^2}.
\]
Thus, a single mode approximation starts working again for short (point-like) JJ and any $\kappa$. Note, that participation of higher modes in the escape makes the escape rate higher, \ie, the barrier height calculated using a single-mode formula lower, exactly as observed in Fig.~\ref{Fig:experiment annular}.

The corresponding barrier height of a point-like JJ is also depicted in Fig.~\ref{Fig:experiment annular}: the dashed line indicates $u_0$ for point-like junctions with a critical current $\gamma_c(\kappa)I_{{\rm c0}}$, given by Eq.~\eqref{Eq:depinning current}. If the only effect caused by the $\kappa$ vortex is the suppression of the critical current of the junction, the experimental data should coincide with the dashed line. However, the experimental data and simulations deviate from this curve, which becomes more pronounced with increasing junction length. This indicates a difference in the escape process of a point-like junction and a long junction with a $\kappa$ vortex.

The white line in Fig.~\ref{Fig:experiment annular} corresponds to the analytic result from Sec.~\ref{Sec:long junctions with a fractional vortex} for infinitely long JJs. Comparing this result to simulations and experimental data, three regimes can be distinguished: a homogeneous regime (small $|\kappa|$, light gray region in Fig.~\ref{Fig:experiment annular}), an intermediate regime (moderate $|\kappa|$, gray region in Fig.~\ref{Fig:experiment annular}) and a vortex escape regime (large $|\kappa|$, dark gray region in Fig.~\ref{Fig:experiment annular}). In the homogeneous regime, the escape process corresponds to the one of a point-like junction: the phase string escapes as a whole from the meta-stable state. The effective barrier height corresponds to $u_0$ in this regime. In the vortex escape regime, the escape-process is dominated by the vortex itself, and experimental data, simulations and theory coincide.
\subsection*{Temperature dependence}
In the thermal regime, the effective barrier height $u$ does not depend on the bath temperature $T$, see Eq.~\eqref{Eq:approximated barrier height}. For temperatures well above the crossover temperature to the quantum regime \cite{Haenggi90}, which for our samples is $\sim 100\,{\rm mK}$, the $u(\kappa)$-dependencies measured at different temperatures should coincide. For lower bath temperatures, $T\ll4.2\,{\rm K}$, the damping dependent prefactor $a$ in Eq.~\eqref{Eq:escaperate} becomes important \cite{Silvestrini88a}. To account for this unknown factor, we fit the escape rate $\Gamma(I)$ calculated from the measured histogram not only with the barrier height, but also with the junctions's resistance $R$ assuming moderate to low damping \cite{Buettiker83} with
\begin{equation}
a=\frac{4}{\left(\sqrt{1+(Qk_B T/1.8u)}+1\right)^2}
  , \label{Eq:damping factor a}
\end{equation}
where $Q=\omega_0 RC$ is the quality factor.
Using this method, the uncertainty of the experimentally obtained $u(\kappa)$-dependence increases.
\begin{figure}[!htb]
  \begin{center}
    \includegraphics{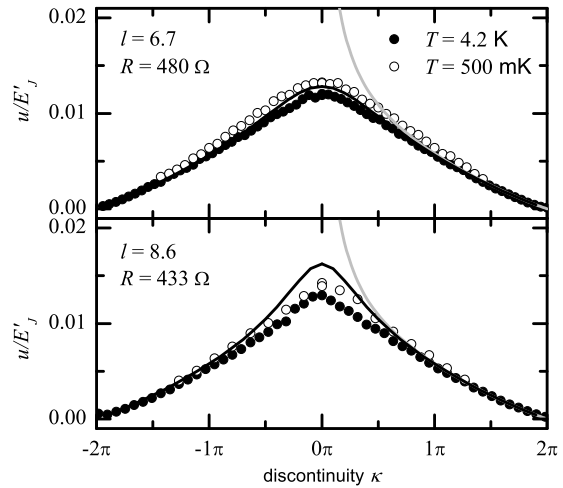}
  \end{center}
  \caption{Effective barrier height $u(\kappa)$ for two different bath temperatures $T$ in the thermal regime. Symbols indicate experimental results, gray lines show theoretical predictions for the vortex escape and solid lines show numerical simulations.
  }
  \label{Fig:T_dependence}
\end{figure}
Fig.~\ref{Fig:T_dependence} compares the barrier height determined at $T=500\,{\rm mK}$ for two annular JJs of lengths $l=6.7$ and $l=8.6$ with the already shown results at $T=4.2\,{\rm K}$, \cf Fig.~\ref{Fig:experiment annular}. The resistance values obtained from best fit are $R=480\,\Omega$ and $R=433\,\Omega$, respectively, corresponding to $Q$-values of $\sim $2000\,...\,3000. Note, the escape rate $\Gamma$ only weakly depends on $R$, which leads to relatively large uncertainties in the determined $R$ values ($\sim 50\%$). Though the method of determining $u$ is less accurate for lower temperatures, the measured $u(\kappa)$-dependence shows very good agreement with numerical simulations.
\section{Linear junctions}
 \label{Sec:Linear junctions}
To study the effect caused by the junction geometry we have also measured linear LJJs with different lengths. As shown in Ref.~\cite{Gaber05}, the $\gamma_c(\kappa)$ pattern of linear junctions differs from Eq.~\eqref{Eq:depinning current} due to the open ends. It is $2\pi$-periodic, has minima at $\kappa=\pi+2\pi n$ and depends on junction length. 
Calibration of the injectors follows the same method as described for annular junctions, however the first minimum of the $I_c(I_{{\rm inj}})$ pattern corresponds here to $\kappa=\pi$. Knowing the values $I^{{\rm min}}_{{\rm inj}}$, the corresponding $\kappa$ can be calculated as $\kappa=\pi I_{{\rm inj}}/\left|I^{{\rm min}}_{{\rm inj}}\right|$ for any $I_{{\rm inj}}$. Note, in contrast to annular junctions, $\gamma_c(\kappa)$ depends on lengths $l$ and has to be determined numerically. In addition, only vortices with $|\kappa|\leq\pi$ can be investigated with the experimental technique described above. For further details see Ref.~\cite{Gaber05}.

\begin{figure}[!htb]
  \begin{center}
    \includegraphics{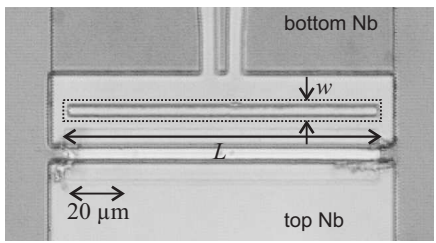}
  \end{center}
  \caption{Optical image of one of the investigated LJJs with one pair of injectors.
  }
  \label{Fig:sample_LJJ}
\end{figure}

For the experiments, we used Nb-Al-AlO$_{x}$-Nb linear LJJs with different lengths, as shown in Fig.~\ref{Fig:sample_LJJ}. The junction properties are listed in Tab.~\ref{Tab:LJJ}.
\begin{table}[!htb]
\centering
\begin{tabular}{|l|l|l|l|l|l|}
\hline
\multicolumn{1}{|c|}{$L$} & \multicolumn{1}{c|}{$w$} & \multicolumn{1}{c|}{$j_c$} & \multicolumn{1}{c|}{$\lambda_J$} &      \multicolumn{1}{c|}{$l$} & \multicolumn{1}{c|}{$I^{{\rm min}}_{{\rm inj}}$} \\ 
\multicolumn{1}{|c|}{[$\mu$m]} & \multicolumn{1}{c|}{[$\mu$m]} & \multicolumn{1}{c|}{[A/cm$^2$]} & \multicolumn{1}{c|}{[$\mu$m]} & \multicolumn{1}{c|}{} & \multicolumn{1}{c|}{[mA]} \\ 
\hline
\multicolumn{1}{|c|}{120} & \multicolumn{1}{c|}{5} & \multicolumn{1}{c|}{84.2} & \multicolumn{1}{c|}{53.5} & \multicolumn{1}{c|}{2.3} & \multicolumn{1}{c|}{4.9} \\ 
\hline
\multicolumn{1}{|c|}{240} & \multicolumn{1}{c|}{5} & \multicolumn{1}{c|}{78.5} & \multicolumn{1}{c|}{55.4} & \multicolumn{1}{c|}{4.4} & \multicolumn{1}{c|}{5.0} \\ 
\hline
\end{tabular}
\caption{Junction parameters (linear) at $T=4.2\,{\rm K}$.}
\label{Tab:LJJ}
\end{table}
\begin{figure}[!htb]
  \begin{center}
    \includegraphics{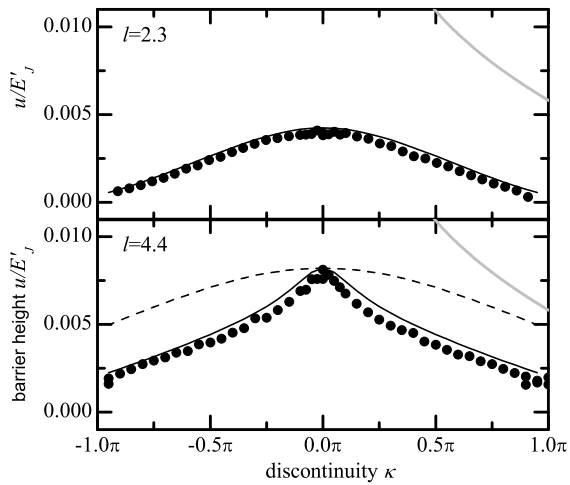}
  \end{center}
  \caption{%
    Effective barrier height as a function of $\kappa$ for linear junctions with different lengths at $T=4.2\,{\rm K}$. Solid circles show experimental results, gray lines show the theoretical prediction and solid lines show numerical simulations. The dashed line indicates the barrier height of an annular junction of the same junction length.
  }
  \label{Fig:experiment linear}
\end{figure}

In Fig.~\ref{Fig:experiment linear} the results for linear LJJs are shown for $T=4.2\,\mathrm{K}$. \emph{Experimental data} and \emph{numerical simulations} show a very good agreement. On the other hand, a comparison of \emph{experimental data} and \emph{theory} shows, that even in the case of the longest junction with $l=4.4$ the experimental barrier height is considerably lower than the asymptotic value $u_v(\kappa)$ even for high values of $\kappa$. Note that for annular junctions all three (experimental data, numerical simulations and analytic calculations) are in good agreement. This discrepancy is due to the junction geometry: the escape in linear junctions is usually dominated by the nucleation of a fluxon-antifluxon pair at the edges. Only at very large $l$ a crossover to vortex activation takes place. At the same time, the observed $\kappa$-dependence suggests that boundary dynamics are affected by the presence of a fractional vortex.
This becomes clear, if one compares numerically simulated $u(\kappa)$-dependencies for two junctions with different geometry, as shown for the longest junction with $l=4.4$: the dashed line in Fig.~\ref{Fig:experiment linear} corresponds to an annular junction of the same length. The obtained barrier heights differ in shape and absolute value.
\begin{figure}[!htb]
  \begin{center}
    \includegraphics{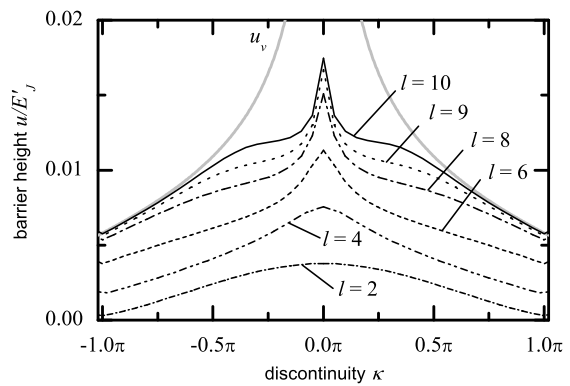}
  \end{center}
  \caption{%
    Numerically calculated effective barrier heights at $\gamma=0.99\gamma_c$ as a function of $\kappa$ for linear junctions with different lengths. The gray line corresponds to analytical results.
  }
  \label{Fig:Simu linear}
\end{figure}
Fig.~\ref{Fig:Simu linear} compares simulated barrier height dependencies on $\kappa$ of linear junctions with different lengths to the theoretical prediction $u_v(\kappa)$ for infinitely long JJs. Only in rather long junctions $(l>9)$ and for high values of $\kappa$ the escape mechanism caused by the fractional vortex becomes important. For $l<9$ and small $\kappa$ the edge effects dominate the phase escape.
\section{Conclusion}
 \label{Sec:Conclusion}
We have experimentally investigated the thermal escape of an arbitrary fractional Josephson vortex close to its depinning current $\gamma_c(\kappa)$ in long JJs of different geometry. A comparison of our results with numerical calculations based on the perturbed sine-Gordon equation shows very good agreement. Depending on the junction's geometry, our results coincide with theoretical predictions in different regimes of $\kappa$, except in the limit $\kappa\rightarrow0$, $l\rightarrow\infty$, where our model, which is based on a single mode approximation, fails \cite{Vogel08}.
\subsection*{Acknowledgements}
K. Buckenmaier gratefully acknowledges support from the Evangelisches Studienwerk e.V. Villigst. This work was supported by the Deutsche Forschungsgemeinschaft via the SFB/TRR21.
\bibliography{UtasBib,datenbank,books}

\end{document}